\documentclass[twocolumn,aps,superscriptaddress,prl]{revtex4}
 
\usepackage{amsthm,amssymb,amsmath}
\usepackage{graphicx}
\usepackage{dcolumn}
\usepackage{bm}
\usepackage{color} %for change tracking

\usepackage{graphicx,graphics}
\usepackage{dcolumn}                % Align table columns on decimal point
\usepackage{amssymb}

\newcommand{\beqa}{\begin{eqnarray}}
\newcommand{\eeqa}{\end{eqnarray}}
\newcommand{\beq}{\begin{equation}}
\newcommand{\eeq}{\end{equation}}

\newcommand{\ket}[1]{| #1 \rangle}
\newcommand{\bra}[1]{\langle #1 |}
\newcommand{\braket}[2]{\left\langle #1 \mid #2 \right\rangle}
\newcommand{\proj}[1]{\left | #1 \rangle \langle #1   \right | }

\newcommand{\tr}{\mathrm{Tr}}

\begin{document}

\title{Sensitivity to perturbations  and  quantum phase transitions}

\author{D. A. Wisniacki and A. J. Roncaglia}
\affiliation{Departamento de F\'isica, FCEyN, UBA and IFIBA, CONICET, 
Pabell\'on 1, Ciudad Universitaria, 1428 Buenos Aires, Argentina}

%\pacs{05.45.Mt}{Quantum chaos; semiclassical methods}
%\pacs{64.70.Tg}{Quantum phase transitions}
%\pacs{05.70.JK}{Critical point phenomena}

%%%%%%%%%%%%%%%%%%%%%%%%%%%%%%%%%%%%%%%%%%%%%%%%%%%%%%%%%%%%%%%%%%%%%%%%%%%%%%%%%%%%%%%%%%%%%%%%%%%%%%%%%%%%%%%%
%%%%%%%%%%%%%%%%%%%%%%%%%%%%%%%%%%%%%%%%%%%%%%%%%%%%%%%%%%%%%%%%%%%%%%%%%%%%%%%%%%%%%%%%%%%%%%%%%%%%%%%%%%%%%%%%
\begin{abstract}
The local density of states or its Fourier transform, usually called fidelity amplitude,
are important measures of quantum irreversibility due to imperfect evolution. 
In this letter we study both quantities in  a paradigmatic many body system: the Dicke Hamiltonian, 
where a single-mode bosonic field interacts with an ensemble of $N$ two-level atoms. 
This model exhibits a quantum phase transition in the thermodynamic
limit while for finite instances the system undergoes a transition 
from quasi-integrability to quantum chaotic.
We show that the width of the local density of states %computed in a region of the spectra 
clearly points out the imprints of the transition from integrability
to chaos but no trace remains of the quantum phase transition. The connection
with the decay of the fidelity amplitude is also established.
\end{abstract}

\maketitle
%%%%%%%%%%%%%%%%%%%%%%%%%%%%%%%%%%%%%%%%%%%%%%%%%%%%%%%%%%%%%%%%%%%%%%%%%%%%%%%%%%%%%%%%%%%%%%%%%%%%%%%%%%%%%
%%%%%%%%%%%%%%%%%%%%%%%%%%%%%%%%%%%%%%%%%%%%%%%%%%%%%%%%%%%%%%%%%%%%%%%%%%%%%%%%%%%%%%%%%%%%%%%%%%%%%%%%%%%%%

%\noindent\textit{Introduction:}
The sensitivity to perturbations is one of the major impediments to fully
control quantum systems.  
With the advent of quantum information and its technological development, which enable the
manipulation  of many
body systems such as cold atoms in optical lattices \cite{Bloch},
a deep understanding of the sources that perturb and deteriorate quantum evolutions 
is required \cite{qinfo,casati}. This would help us to develop strategies to protect 
and manipulate quantum systems, but also by analysing the response to 
perturbations one would be able to extract information from the actual dynamics.
 
In quantum evolutions, the effects of perturbations can be analyzed by 
measuring how difficult is to reverse a given dynamics, as it was proposed by Peres \cite{peres}.
To this end several figures of merit have been defined. Among them, the so-called 
local density of states (LDOS) or strength function, defined by Wigner \cite{wigner} to
describe  the statistical behavior of perturbed eigenfunctions, has been extensively studied 
due to its connections with fundamental problems such as irreversibility, thermalization or 
dissipation in quantum systems \cite{rigol, santos}. 
Moreover, the LDOS provides significant information in 
quantum quenches, one of the simplest nonequilibrium quantum phenomena \cite{silva}.
Consider a one parameter dependent Hamiltonian $H(\lambda)$,
with eigenenergies $E_j (\lambda)$ and eigenstates $\ket{j(\lambda)}$.
The LDOS of an eigenstate $\ket{i(\lambda_0)}$, that we call unperturbed,  is defined as
\beq
\rho_i(E,\delta \lambda)=\sum_j|\braket{j(\lambda)}{i(\lambda_0)}|^2 \delta(E-E_{ij}),
\label{ldos-eq1}
\eeq
where $\delta \lambda=\lambda-\lambda_0$ and $E_{ij}=E_j(\lambda)-E_i(\lambda_0)$.
It is the distribution of the overlaps squared between the unperturbed and perturbed 
eigenstates. The LDOS has been studied in several systems with different perturbations 
\cite{wigner,Flambaum,Fyodorov,Casati,Cohen,Wisniacki}, and it is equivalent 
to the probability of work for a quantum quench \cite{silva}.
This quantity is also intimately related to other measures of 
irreversibility. 
In fact, the averaged LDOS is equal to the Fourier transform of the fidelity amplitude  (FA),
\beq
O(t)=\tr[U^\dagger_{\lambda_0+\delta\lambda}(t) U_{\lambda_0}(t)]
\eeq 
where $U_{\lambda_0}(t)$ is the evolution operator corresponding to the Hamiltonian
$H(\lambda_0)$, and  $U_{\lambda_0+\delta\lambda}(t)$
corresponds the perturbed one that governs the  backward evolution. 
Further, $O(t)$ is connected with other well-known 
quantity, the Loschmidt echo (LE), defined as  
$M_{\psi}(t)=|\bra{\psi} U^\dagger_{\lambda_0+\delta\lambda} (t) U_{\lambda_0} (t)\ket{\psi}|^2$
for a given initial state $\ket{\psi}$. If we average the LE over 
initial states according to Haar measure (which is uniform over all quantum states in the Hilbert space) we obtain \cite{zanardi04,dankert}:
\beqa
\overline{M}(t)= \int d\ket{\psi} \, M_{\psi}(t) =
\frac{[d+|O(t)|^2]}{d(d+1)},
\eeqa
where $d$ is the dimension of the Hilbert space. Thus, the 
width of the LDOS gives the characteristic time-scale for the
decay of the FA and the averaged LE.

During the last years, a great deal of work has been devoted to characterize  
the sensitivity to perturbations and irreversibility using these three quantities,
the LDOS, FA or LE \cite{prosen-rev,jacquod,scholarpedia}.  Several regimes were
shown, and some of them appear to be universal \cite{jalabert, diego}. 
Despite the importance of correlated many body systems, not only from a theoretical point of view
but also in actual experimental setups, most of these studies were focused 
in single body systems. Only a few recent contributions consider the LE and the FA
for many body systems \cite{Izrailev,zanardi,zanardi2,zanardi3,OpSuscep,manfredi1,manfredi2,Pastawski}. 
In Ref.  \cite{zanardi,zanardi2,zanardi3,OpSuscep} it is shown 
that the LE of the ground state is  a good indicator of a quantum phase transition.  
These studies were carried out for a one-dimensional transverse Ising model and 
a Heisenberg spin chain by considering the ground state fidelity. 
Other works that consider the evolution of a many body system,  
approximated by self consistent hydrodynamical equations, found that the LE 
drops abruptly after a critical time \cite{manfredi1,manfredi2}. 
Despite the above evidence, little is known about the behaviour of the FA and the 
LDOS for general evolutions, where the excited region of the spectra of such many body 
systems is involved.

The goal of this letter is to study the LDOS and 
the FA in a paradigmatic many body system: the Dicke model, where
a single bosonic field interacts with $N$ two-level atoms.
This model exhibits a quantum phase transition in the 
thermodynamic limit ($N\rightarrow \infty$)
when the parameter $\lambda$, that controls the strength of the interaction, 
crosses a critical value $\lambda_c$.
On top of that, for finite $N$, the system undergoes a transition 
from quasi-integrability to quantum chaotic within the same region of parameters. 
Remarkably, an experimental realization of this model has been recently done 
using a superfluid gas in an optical cavity \cite{nature-dicke}.

Here we show that the width $\Gamma$ of the 
LDOS, which provides the time-scale for the decay of the FA, has a well defined
behaviour depending on which side of the transition belongs the
unperturbed evolution. In the case where  $\lambda_0<\lambda_c$, the width of the LDOS
is a linear function of the strength of the perturbation $\delta \lambda$. 
However, if $\lambda_0>\lambda_c$,  three regimes are observed. For sufficiently small
$\delta \lambda$, so that first order perturbation theory is valid, the width grows
linearly $\Gamma \sim  \delta \lambda$. Then, a crossover to a Fermi
golden rule regime in which $\Gamma \sim  \delta \lambda^2$ is
observed. Finally, for larger perturbations, $\Gamma$ grows  linearly again. 
 These results are consistent with those  obtained in a banded random model 
initially studied by Wigner \cite{wigner, Casati}. In order determine whether the source
of this behaviour is due to presence of the quantum phase transition, we also considered 
the Dicke model in the rotating wave approximation, where the 
Hamiltonian  is quasi-integrable for every $\lambda$, but also displays a quantum phase 
transition. By comparing the results in these two situations, we were able to show
that the transition in the behaviour of $\Gamma$  
is related to the integrability-chaos transition and not to the quantum phase transition.
Finally, we consider the decay of the FA, $O(t)$, and show the relation between the first 
two regimes of $\Gamma$ and the decay of $O(t)$. 

%\noindent\textit{The model.--}
We begin by describing the system that we consider: 
the single-mode Dicke Model. This model describes 
an ensemble of $N$ two-level atoms with level  splitting $\omega_0$ 
coupled to a single bosonic mode of frequency $\omega$ via dipole interaction  ($\hbar=1$):
\beq
  H(\lambda) 
 = \omega_0 J_z+\omega\, a^\dagger a+\frac{\lambda}{\sqrt{2j}}(a^\dagger + a)(J_+ + J_-),
 \label{eq:DickeH}
\eeq
in this case $J_z$ and $J_{\pm}$ are the collective angular momentum operators 
for a pseudospin of length $j=N/2$, 
$a$ and $a^\dagger$ are the bosonic operators of the field, and $\lambda$ is the atom-field
coupling constant. 
In the thermodynamic limit, 
$N\rightarrow\infty$, this model exhibits a quantum phase 
transition at $\lambda_c=\sqrt{\omega\omega_0}/2$ where there is broken 
symmetry associated to the parity \cite{Hepp73}.
When $\lambda<\lambda_c$ the system is in the normal phase, while
for $\lambda>\lambda_c$ the system is in the superrandiant phase.
 For finite sized instances
and sufficiently high $N$ it displays a 
crossover in its level statistics from Poissonian to a Wigner distribution 
at $\lambda\approx\lambda_c$  \cite{Brandes}.
We shall call this transition from quasi-integrable to quantum chaos. 
In this case the parity, $\Pi=\exp(i\pi\hat N)$ with 
$\hat N=a^\dagger a+J_z+j$ the so-called excitation number, 
is a conserved quantity. Thus, the Hilbert space  is split into two 
non-interacting subspaces with definite parity. 
%In the limit of $\lambda\rightarrow\infty$ the system becomes regular \cite{Brandes}. 
On the other hand, if we neglect the counter rotating terms in the interaction, 
by applying the rotating wave approximation (RWA) \cite{Haroche}, we can then define the 
following Hamiltonian:
\beq
H_{RWA}(\lambda)=\omega_0 J_z +\omega a^\dagger a
+{\frac{\lambda}{\sqrt{2j}}}(a^\dagger J_-+aJ_+),
\label{eq:RWAH}
\eeq
that also exhibits a quantum phase transition in the thermodynamic limit, but it is
quasi-integrable for every finite $N$ and $\lambda$ \cite{Brandes}.  

We will consider the system away from the thermodynamic limit. 
Since the parity is a conserved quantity, we have restricted 
to the odd subspace. The parameters that were used in our numerical simulations are such that 
the system is in scaled resonance, $\omega=\omega_0=1$, so that $\lambda_c=0.5$ and 
$\lambda_c^{RWA}=1$;
$j=20$ corresponding to $N=40$; and we have truncated the bosonic mode to $n=350$. 
For the Hamiltonian of eq. \eqref{eq:DickeH} we have also checked that $N$ is high enough so 
that the level statistics  obeys a Wigner distribution for $\lambda > \lambda_c$.
Since we truncate the Hilbert space of the bosonic
mode, we consider excited states whose energy does not change
as the value of $n$ is increased. % in the numerical simulations. 
This was done in order to avoid numerical errors due to the truncation.
In order to smooth fluctuations arising from individual wave functions, 
we compute the averaged LDOS: $\bar{\rho}(E,\delta \lambda)$.
This is equivalent to consider a generalized LDOS for a microcanonical state located in
a given energy window, $\frac{1}{N_\Delta}\sum_{j, |E-E_j| < \Delta}\proj{j(\lambda_0)}$.
The average was done using a window of 200 states
around the eigenstate 500, similar results were obtained by 
averaging over other energy windows.
In order to determine $\Gamma$ we have considered the distance 
from the mean value of the LDOS that contains 70$\%$ of the probability.
That is, $
  \int^{\langle E \rangle + \sigma}_{\langle E \rangle - \sigma} \rho(E, \delta \lambda) dE = 0.7 \hspace{0.4cm} 
$
where $\langle E \rangle= \int E \; \rho(E, \delta k) d E $. 
Remarkably, the width of the averaged LDOS has another interesting interpretation
as the fluctuations in the probability of work for a quantum quench starting from 
a microcanonical state in given energy window \cite{silva}.

\begin{figure}[h]
\begin{center}
\includegraphics[width=0.8\linewidth]{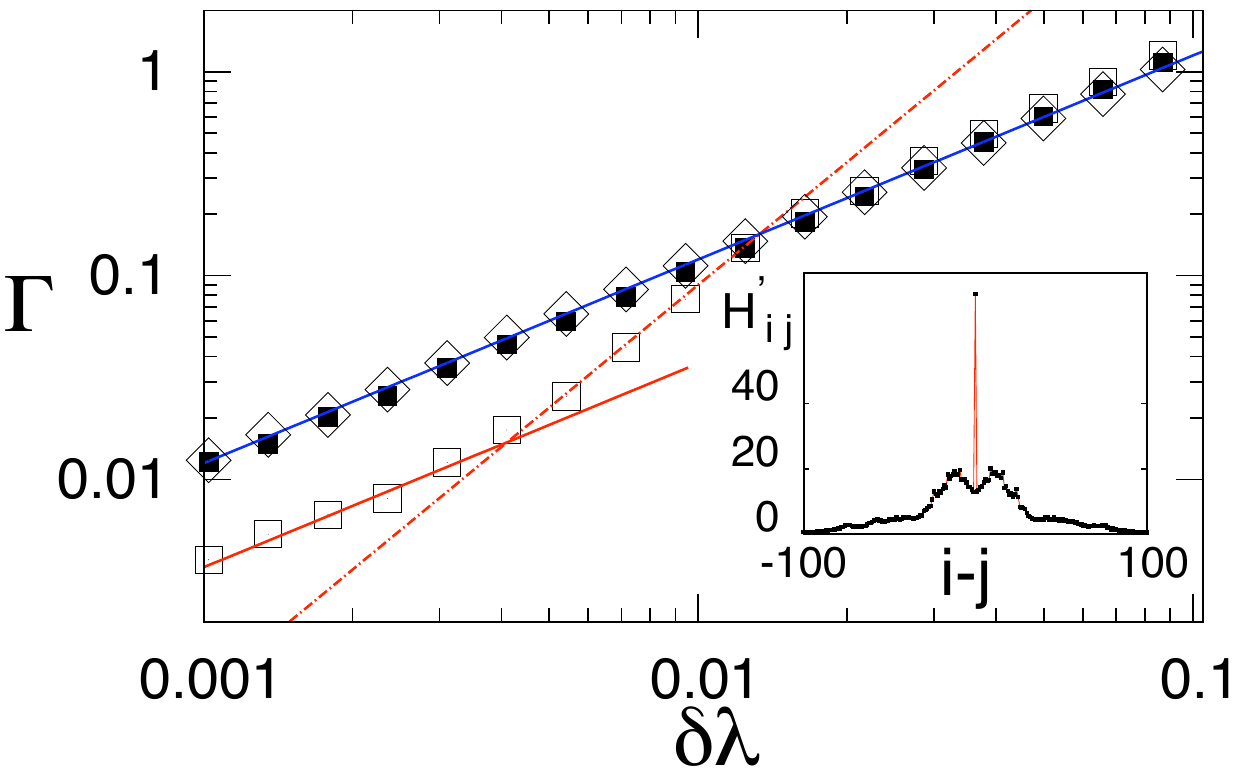}
\caption{(Color online) Width of the LDOS as a function of the perturbation $\delta\lambda$
for: $\blacksquare$ $\lambda_0=0.1$; $\square$ $\lambda_0=0.8$; $\diamond$
$\lambda_0 = 1.6$ with the RWA.  Inset: Mean value of the matrix elements $|H^{'}_{i \; j}|$  as a function of $i-j$ for $\lambda_0=0.8$ in the unperturbed basis. See text for details.}
\label{LDOS-1}
\end{center}
\end{figure}

In Fig. \ref{LDOS-1} we display the width of the LDOS, $\Gamma$, 
as a function of the perturbation for some values of
$\lambda_0$. 
There we can observe different behaviors depending on whether the Hamiltonian is
quasi-integrable. On one hand, we can see a linear dependence with the perturbation
for quasi-integrable systems, 
i.e. $H(\lambda_0)$ (with small $\lambda_0$) and $H_{RWA}(\lambda_0)$.
On the other hand, for $H(\lambda_0)$ and $\lambda_0 > 0.5$, we can identify three
different regimes as a function of $\delta\lambda$. For small
perturbations $\Gamma$ is a linear function of $\delta\lambda$, 
for moderate values of $\delta \lambda$  there is quadratic dependence,
and finally for strong perturbations a linear dependence is
achieved. 
The initial linear regime corresponds to the 
situation where first order perturbation holds. In this case, the matrix elements of the
perturbation, defined as $H' \equiv [H(\lambda_0+\delta \lambda)-H(\lambda_0)]/\delta \lambda$,
are such that
$\frac{|H' _{i,j}| \delta \lambda}{\Delta E}<1,$
where $\Delta E$ is the mean level spacing. In the inset of Fig.~\ref{LDOS-1} we plot the value of the matrix elements of the perturbation
in the basis of the unperturbed eigenstates.
In the region of the spectra that we considered, $\Delta E\approx~0.07$ so that $\delta\lambda<0.003$.  Thus, for $\delta\lambda>0.003$ a 
crossover from a linear to a quadratic regime is observed. 

\begin{figure}[h]
\begin{center}
\includegraphics[width=0.8\linewidth]{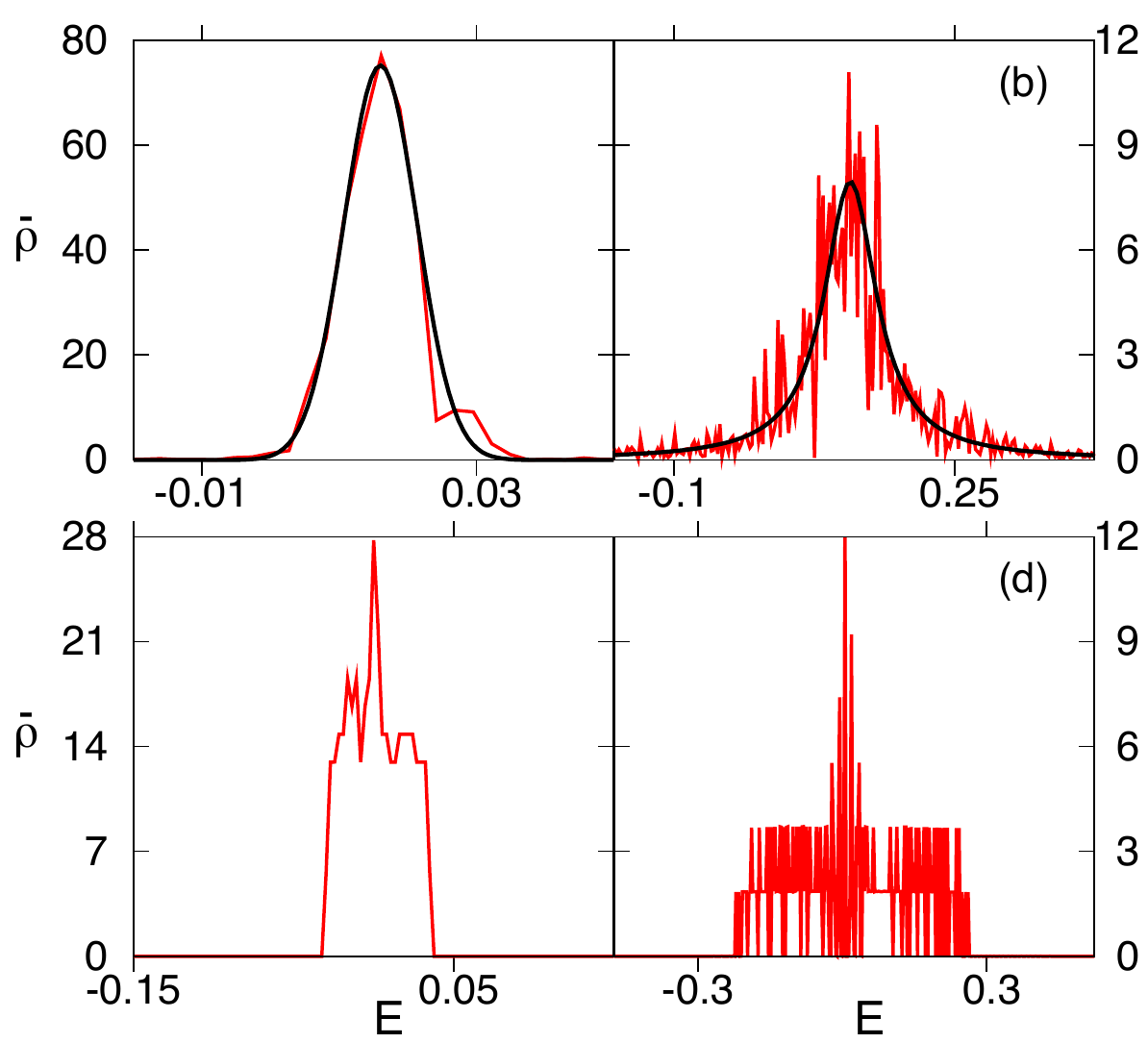}
\caption{Mean LDOS $\bar{\rho}(E,\delta \lambda)$ for the Dicke model. The average was done using 200 states around
the eigenstate with energy $E_{500}$. (a) $\delta \lambda=0.001$ and $\lambda_0=0.8$, (b) $\delta \lambda=0.08$ and $\lambda_0=0.8$, (c) $\delta \lambda=0.001$ and $\lambda_0=0.2$ and (d) $\delta \lambda=0.08$ and $\lambda_0=0.2$.
}
\label{meanldos}
\end{center}
\end{figure}

\begin{figure}[h]
\begin{center}
\includegraphics[width=0.8\linewidth]{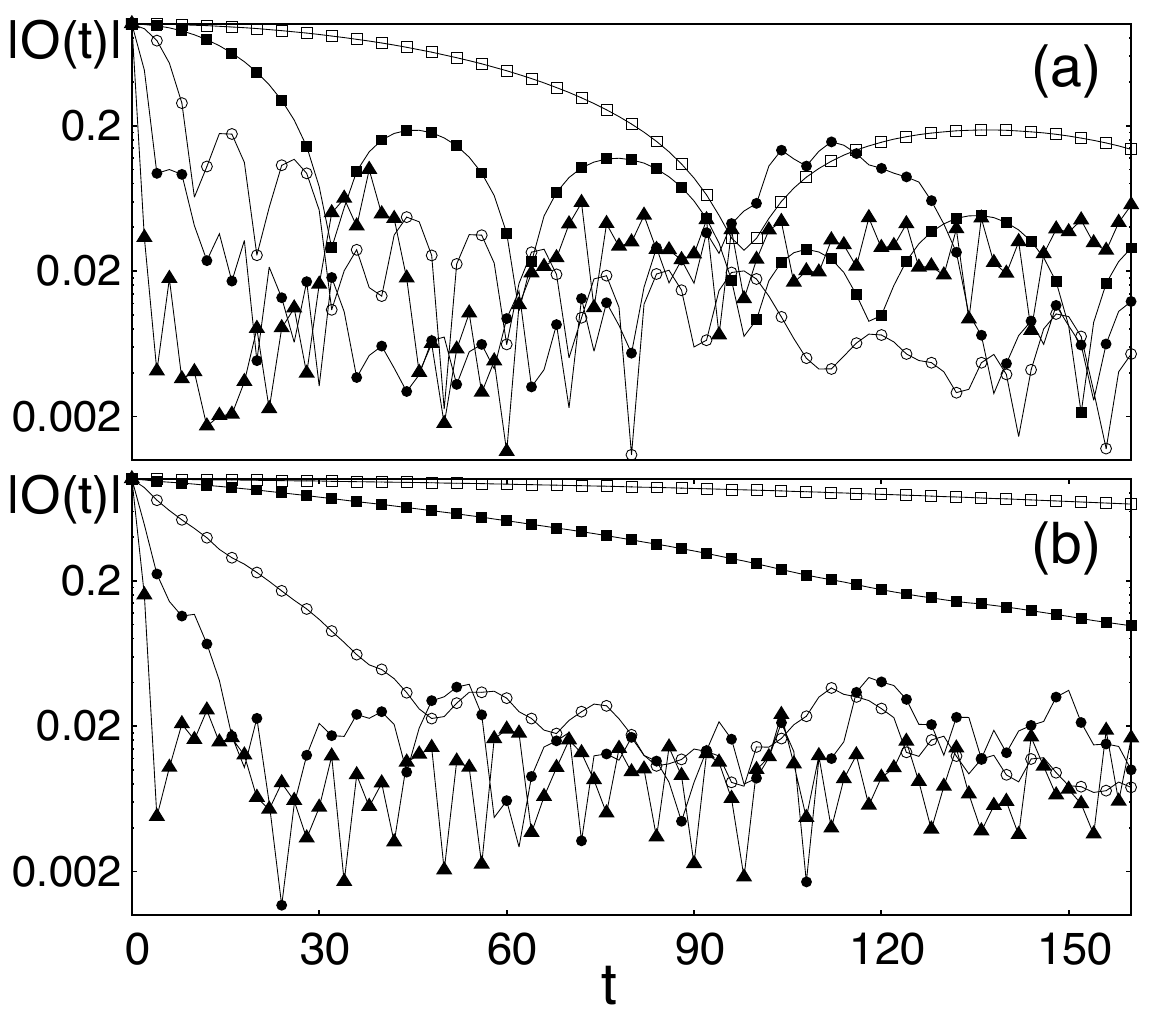}
\caption{Modulus of the FA $|O(t)|$ as a function of time. In (a) $\lambda=0.1$ and
(b) $\lambda=0.8$, both without the RWA. The FA was computed using 200 states around
the eigenstate with energy $E_{500}$.
$\square$ $ \delta \lambda= 10^{-3}$,
$\blacksquare$ $\delta \lambda=3.1\, 10^{-3}$,
$\circ$ $\delta \lambda=9.4\, 10^{-3}$,
$\bullet$ $\delta \lambda=2.9\, 10^{-2}$, $\blacktriangle$ $\delta \lambda=8.7\, 10^{-2}$.}
\label{Fidelity}
\end{center}
\end{figure}

As we will see, the appearance of a quadratic
regime determines the range of perturbations where 
the LDOS has a Lorentzian shape [see Fig.~\ref{meanldos} (b)]. 
Finally, for strong perturbations, a regime where $\Gamma$ depends linearly on $\delta \lambda$
is achieved, this is the non-perturbative regime. The last two regimes have been  
observed also in the banded random matrix model, as the one studied by Wigner \cite{wigner,doron}. Another conclusion that we can extract is that, if one 
considers moderate values of perturbations, the
decay of the FA, given by the width of the LDOS, is faster for the quasi-integrable 
Hamiltonian than for the chaotic one. This is in agreement with the numerical 
simulations that are shown bellow [see Fig.~\ref{Fidelity}].

Let us now consider the structure of the LDOS. In Fig. \ref{meanldos}
we show the typical behavior of the mean LDOS for the Dicke
model. If we consider $\lambda=0.8$ for $\delta\lambda <0.003$, as we discussed before, the LDOS
is a Gaussian distribution [see Fig. \ref{meanldos} (a)]. This regime
is characterized by the validity of the first order perturbation
theory, so the overlap of unperturbed and perturbed states is
approximately $\braket{j(\lambda)}{i(\lambda_0)}\sim \delta_{i j}$ 
and the gaussian distribution comes from  the
distributions of the eigenenergies that appear in Eq.
\ref{ldos-eq1}. For greater values of $\delta \lambda$, first order perturbation
theory is no longer valid and  a crossover to a
Lorentzian distribution is observed [see Fig. \ref{meanldos} (b)]. 
In Fig. \ref{meanldos} (c) and (d) we show the corresponding 
mean LDOS when $\lambda<\lambda_c$. In this case, 
the LDOS has no recognizable structure. Similar distributions
were obtained for the RWA Hamiltonian. 

\par
\begin{figure}[h]
\begin{center}
\includegraphics[width=0.8\linewidth]{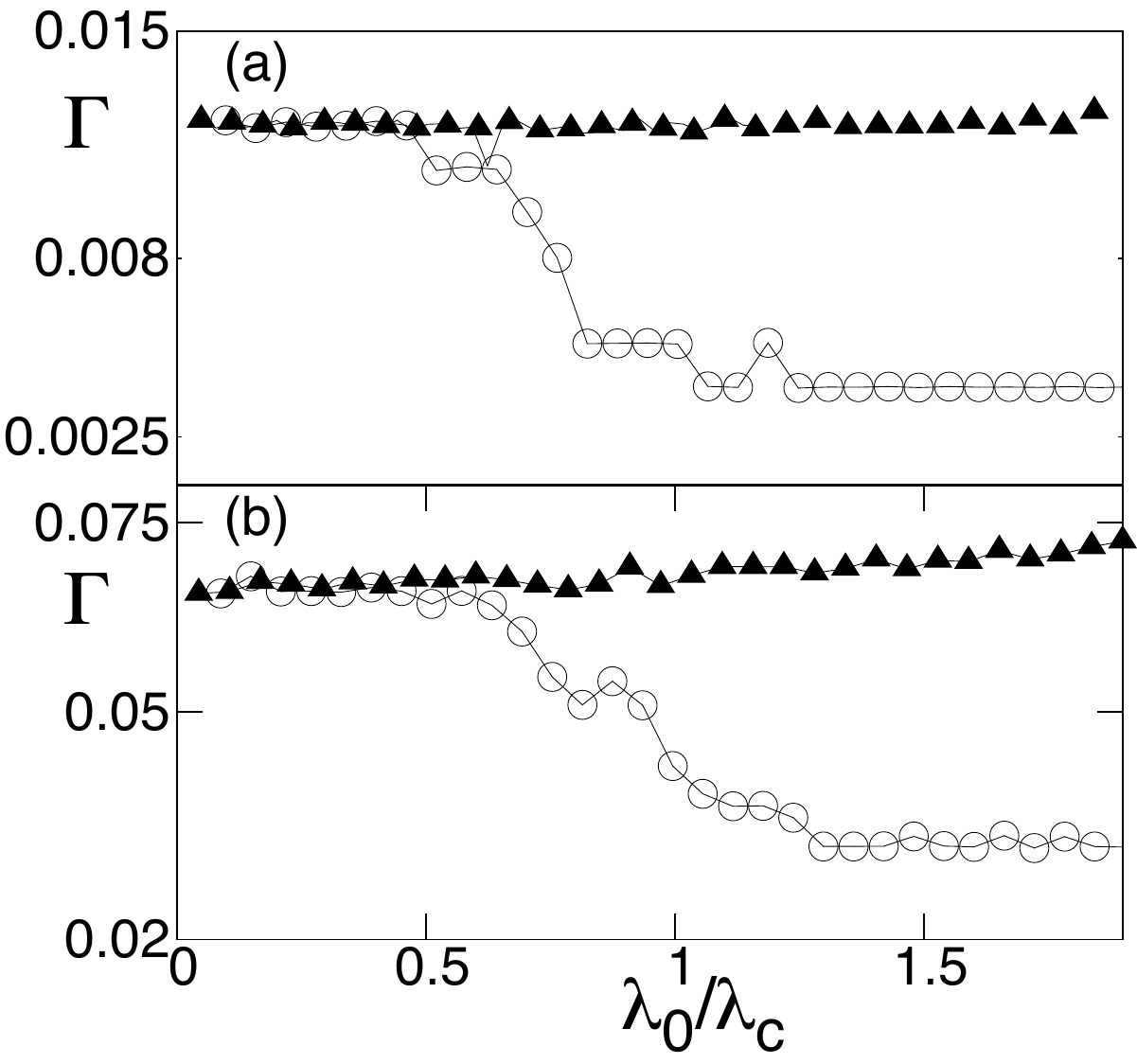}
\caption{Width of the LDOS, $\Gamma$, as a function of $\lambda_0$ for a fixed small perturbation $\delta\lambda$.
$\blacktriangle$ RWA, $\circ$ without RWA. In (a) $\delta \lambda=0.001$ and (b) $\delta \lambda =0.06$.}
\label{LDOS-Lamb}
\end{center}
\end{figure}

 The transition from quasi-integrability to quantum chaos that appears in the Dicke
Hamiltonian is reflected in the behavior of the width of the LDOS. 
This becomes evident when we compare the above results with the case where no such transition is
present, the RWA Hamiltonian. In addition to this change in the spectral statistics, 
there is a quantum phase transition in the thermodynamic limit in both systems. 
However, Fig. \ref{LDOS-1} seems to indicate that no trace of this transition 
is present in the width of the LDOS for an excited region of the spectra. In order to show this 
in more detail, in Fig. \ref{LDOS-Lamb} we plot the width of the LDOS in 
terms of  $\lambda$ for a fixed small perturbation $\delta \lambda$. 
In Fig. \ref{LDOS-Lamb} (a) we consider  $\delta \lambda=0.001$, so we 
are in the regime where $\Gamma$ depends linearly with $\delta \lambda$ for the Dicke Hamiltonian
and its RWA. In Fig. \ref{LDOS-Lamb} (b) $\delta \lambda=0.06$, so  $\Gamma$ depends
quadratically with $\delta \lambda$ for the Dicke Hamiltonian. 
From the plot we can see that for small enough values of $\lambda$ this function is the same 
for both Hamiltonians, reflecting the fact that the RWA is a good 
approximation for small values of $\lambda$. As $\lambda$ is increased 
the value of $\Gamma$ decreases for the full Hamiltonian up to
a value which is approximately $\lambda_c$ where it remains constant again.
While when we consider the RWA Hamiltonian $\Gamma(\lambda)$ remains 
approximately constant for the full range. Therefore, $\Gamma(\lambda)$ behaves as an
indicator of the quasi-integrable to quantum chaotic transition, but it does not show any trace 
of the quantum phase transition that is also present in the RWA.  
A related quantity that behaves in a similar way is the operator fidelity metric \cite{OpFidelity} but, in contrast the width of the LDOS, it is time-dependent. 

\par We turn now to the discussion of the behavior of the FA. 
In Fig. \ref{Fidelity} we consider the modulus of the FA as a function of time. 
As in the previous results, the FA was computed by averaging 200 states around
the eigenstate 500. We show some examples for the quasi-integrable 
region and for the quantum chaotic region. 
In Fig.\ref{Fidelity} (a) $\lambda=0.1$, so the system is quasi-integrable and in 
(b) we show the chaotic case using $\lambda=0.8$ \cite{Brandes}.
Comparing the decays for the same $\delta \lambda$ of  Fig.\ref{Fidelity}
(a) and (b) we can clearly see that if $\delta \lambda < 0.01$ the 
quasi-integrable case decays faster than the chaotic one.
If  $\delta \lambda > 0.01$  both cases decay approximately in the same way.
As we said above, we could extract the same conclusion from looking at 
the width of the LDOS in Fig. \ref{LDOS-1}, which
provides a characteristic time-scale for the decay of the FA.
Similar behavior was previously observed in one body systems \cite{prosen-rev}.
We would like to remark that, similarly to what happens in 
\cite{Pastawski}, in this many body system no signatures
of hypersensibility in which the FA or LE drops abruptly was observed \cite{manfredi1,manfredi2}.

We have also analyzed the short time decay of the modulus of the FA.
When the system is quasi-integrable [Fig.\ref{Fidelity} (a)]
and for  $\delta \lambda<2 \,  10^{-2}$, the decay at short times is 
essentially Gaussian, and also displays some oscillations due to degeneracy.
But, if the system is chaotic we can show that the time dependence of $|O(t,\delta \lambda)|$ 
is of the form,
\beq
|O(t,\delta \lambda)|\approx a\,e^{-b^2 t^2}+(1-a)\, e^{-c t}.
\eeq
for appropriate $a$, $b$ and $c$ that depend on $\lambda$ and $\delta\lambda$. For small
perturbations is a linear combination of Gaussian and an exponential decay.
As the perturbation is increased the value of $a$ tends to zero, and for the
region where $\Gamma$ is quadratic with $\delta\lambda$ 
(see also Fig. \ref{LDOS-1}) we recover the exponential decay. 

Summarizing,  we have considered the sensitivity to perturbations and
the irreversible dynamics in the critical Dicke model by using the LDOS and the FA.
We have studied the width of the LDOS, which defines the time-scale for the decay of the FA,
and showed the appearance of three different regimes,
depending on the strength of the perturbation, for the chaotic Hamiltonian. 
These regimes were also observed in a banded random matrix model defined by 
Wigner \cite{wigner, Casati}.
On the other hand, for integrable Hamiltonians the width of the LDOS increases linearly with
perturbation. We showed that the decay of the fidelity amplitude, given by the width of the LDOS $\Gamma(\lambda)$, is sensitive to the transition from quasi-integrability to quantum chaos.
However, a proper comparison with its RWA shows that no 
trace of the phase transition can be found in the excited spectra.
Thus, the FA is unable to detect the quantum phase transition unless the ground 
state fidelity is considered. Finally, we would also like to stress
that our results have further applications in relation to the probability of work in 
quantum quenches.

%%%%%%%%%%%%%%%%%%%%%%%%%%%%%%%%%%%%%%%%%%%%%%%%%%%%%%%%
%%%%%%%%%%%%%%%%%%%%%%%%%%%%%%%%%%%%%%%%%%%%%%%%%%%%%%%%
\acknowledgments
The authors acknowledge the support from CONICET (PIP-6137) , UBACyT (X237, 20020100100741, 20020100100483) 
and ANPCyT (1556). We would like to thank Ignacio Garc\'{\i}a Mata for useful discussions.
%%%%%%%%%%%%%%%%%%%%%%%%%%%%%%%%%%%%%%%%%%%%%%%%%%%%%%%%
%%%%%%%%%%%%%%%%%%%%%%%%%%%%%%%%%%%%%%%%%%%%%%%%%%%%%%%%

\end{document}